\documentclass[prb, twocolumn, superscriptaddress, amssymb, floatfix]{revtex4-2}

\usepackage[colorlinks=true]{hyperref}

\usepackage{graphicx}
\usepackage{dcolumn}
\usepackage{bm}
\usepackage[intlimits]{amsmath}
\usepackage{mathtools}
\usepackage{physics}

\usepackage{appendix}
\usepackage{dsfont}
\usepackage[mathscr]{euscript}

\usepackage[caption=false]{subfig} 
\usepackage[capitalize,nameinlink]{cleveref} 

\usepackage[english]{babel}

\DeclareMathOperator{\dt}{dt}

\newcommand{\ham}{\hat{\mathcal{H}}}

\newcommand{\mbf}[1]{\mathbf{#1}}

\newcommand{\mc}[1]{\mathcal{#1}}
\newcommand{\ms}[1]{\mathscr{#1}}
\newcommand{\mt}[1]{\mathrm{#1}}

\newcommand{\expvalp}[1]{\langle{#1}\rangle}

\graphicspath{{Figs/}}

\begin{document}

\title{Spectral functions with infinite projected entangled-pair states}

\author{Juan Diego Arias Espinoza}
\email{j.d.ariasespinoza2@uva.nl}
\affiliation{Institute for Theoretical Physics, Institute of Physics, University of Amsterdam, Science Park 904, 1098 XH Amsterdam, The Netherlands}

\author{Philippe Corboz}
\affiliation{Institute for Theoretical Physics, Institute of Physics, University of Amsterdam, Science Park 904, 1098 XH Amsterdam, The Netherlands}

\begin{abstract}
Infinite projected entangled-pair states (iPEPS) provide a powerful tool to study two-dimensional strongly correlated systems directly in the thermodynamic limit. In this work, we extend the iPEPS toolbox by a method to efficiently evaluate non-equal time two-point correlators, enabling the computation of spectral functions. It is based on an iPEPS ansatz of the ground state in a large unit cell, with an operator applied in the center of the cell, which is evolved in real-time using the fast-full update method. At every time step, the two-point correlators within a cell are computed based on the corner transfer matrix renormalization group method. Benchmark results for the 2D transverse field Ising model show that the main features of the dynamical structure factor can already be reproduced at relatively small bond dimensions and unit cell sizes. The results for the magnon dispersion are found to be in good agreement with previous data obtained with the iPEPS excitation ansatz. 
\end{abstract}

\maketitle

\section{Introduction}
Tensor networks provide very accurate variational ans\"atze for strongly-correlated ground states of local Hamiltonians, where the accuracy is systematically controlled by the bond dimension of the tensors. The best known example is the matrix product state (MPS), the underlying ansatz of the density matrix renormalization group (DMRG) method~\cite{white1992}, which has had an enormous impact in the study of (quasi) 1D systems. Projected entangled-pair states (PEPS)~\cite{verstraete2004,nishio2004} are a higher-dimensional generalization of MPS, enabling the efficient representation of large 2D systems, or even infinite 2D systems,  called infinite PEPS (iPEPS)~\cite{jordan2008}. PEPS and iPEPS have proven to be powerful tools for a broad range of challenging problems, ranging from frustrated magnets to strongly correlated electron systems, not only for ground states, but also at finite temperature, see e.g. Refs.~\cite{corboz14_shastry, liao17, zheng17, niesen17, chen18, lee18, jahromi18, yamaguchi18,kshetrimayum19b, lee20, gauthe20, hasik21, shi22, liu22b, ponsioen23b, weerda23, hasik24, czarnik19c, jimenez21, czarnik21, gauthe22, sinha22}.

Recently, iPEPS has been extended to study also low-energy excitations, based on an excitation ansatz which was originally developed for MPS~\cite{haegeman13,haegeman13b} and later generalized to iPEPS~\cite{vanderstraeten2015}.  It relies on the idea that a  quasiparticle-like excitation can be constructed based on a local perturbation of the ground state, combined with a momentum superposition, such that the ansatz has a well-defined momentum by construction. Several methods to optimize and evaluate the ansatz have been developed and tested for various models~\cite{vanderstraeten19,ponsioen20,ponsioen22,ponsioen23,tu24}. Based on the set of all excitations, it is possible to compute spectral functions (dynamical structure factors)~\cite{ponsioen22, Chi22, ponsioen23, tu24}, enabling a direct comparison with experiments, e.g. from inelastic neutron scattering or angle resolved photo emission spectroscopy. Due to the nature of the excitation ansatz, it can accurately represent spectral features associated with quasiparticle-like excitations, whereas a continuum of excitations can only be resolved in a limited way.

A complementary way to obtain spectral functions is based computing the non-equal time two-point correlators via a real-time evolution.  A common approach is to time evolve the ground state with an operator applied, corresponding to a local quench of the ground state, from which the dynamical correlation functions can be extracted at each time step. This approach has been successfully applied to a broad range of problems using MPS in 1D and 2D systems on cylinders, see e.g. Refs.~\cite{daley04, white04b,zaletel15, gohlke17,verresen18,verresen19,paeckel19}. More recently, results based on isometric tensor networks states~\cite{lin2022} (a subclass of PEPS) or Neural Quantum states~\cite{mendes-santos2023}  have been obtained for 2D systems. A similar scheme based on iPEPS would be desirable. However,  while there have been several works on iPEPS time-evolution methods~\cite{czarnik19, hubig19, kshetrimayum20, dziarmaga2021,kshetrimayum21, dziarmaga2022b,kaneko22,ponnaganti23,kaneko23,dziarmaga23}, these were mostly focused on global quenches. An approach to study local quench dynamics  was introduced in Ref.~\cite{hubig2020b}, based on a translational invariant iPEPS with additional ancilla sites to represent a superposition of local perturbations. However, the time evolution of such a superposition may require a larger bond dimension, due to a larger entanglement, than in a standard local quench.

In this work, we introduce a practical approach to compute spectral functions based on local quench dynamics with iPEPS. The main idea is to start from the iPEPS ground state in a periodically repeated unit cell with an operator applied in its center, which is evolved in real-time. At each time step, the two-point correlators with respect to the center site are evaluated based on the corner transfer matrix renormalization group (CTMRG) method. The unit cell size is kept large enough such that interaction effects between neighboring cells are negligible. We present benchmark results for the transverse-field Ising model for different field strengths, and compare them  with previous data obtained from the iPEPS excitation ansatz. Our results provide a proof of principle that real-time evolution-based methods can be used to study spectral properties of 2D systems with iPEPS.

This paper is organized as follows. We first provide a brief introduction to iPEPS in Sec.~\ref{sec:iPEPS} and summarize the full-update real-time evolution algorithm in  Sec.~\ref{sec:FU}. In Sec.~\ref{sec:DSF}, we introduce the main algorithm of this work, the finite-cell dynamical iPEPS method, to compute spectral functions, where we focus on the dynamical structure factor in spin systems. 
 In Sec.~\ref{sec:results} we present and discuss the benchmark results for the transverse field Ising model, first for the local quench dynamics, followed by the dynamical structure factor data. We analyze the finite bond dimension and finite cell size effects on the results, and discuss the limitations on the maximal simulation times. 
 Finally, in Sec.~\ref{sec:outlook} we summarize or main conclusions and outline possible future directions for further improving the approach.

\section{Method}

\subsection{Introduction to iPEPS}
\label{sec:iPEPS}

An iPEPS is a variational tensor network ansatz to represent 2D wave functions in the thermodynamic limit~\cite{verstraete2004, nishio2004, jordan2008}. It consists of a  of a set of tensors $\left\{ \mbf{A}[\vec{r}] \mid \vec{r} \in \ms{C} \right\}$ in a unit cell $\ms{C}$   which is periodically repeated on the lattice, with  one tensor per lattice site. Each tensor acts as a map from four auxiliary Hilbert spaces of dimension $D$ into a local physical Hilbert space of dimension $d$, with $d=2$ for a spin $S=1/2$ system. $D$ is called the bond dimension and controls the accuracy of the ansatz. For translationally invariant states, a unit cell of size $L=1$ with a only single tensor $\mbf{A}$ can be used.

To calculate an expectation value of some local operator $\hat{O}$ of an iPEPS $\ket{\Psi}$, i.e. $\expvalp{\hat{O}} \coloneqq \bra{\Psi}\hat{O}\ket{\Psi}/\bra{\Psi}\ket{\Psi}$, the corresponding infinite tensor network needs to be contracted in an approximate way, which in this work is done using the CTMRG method~\cite{nishino1996, orus2009, corboz2011, corboz14_tJ}. In CTMRG the infinite tensor network surrounding a central site is effectively encoded in an environment $\ms{E}$ consisting of four corner and four edge tensors. These environment tensors are obtained iteratively by a real-space renormalization type of procedure, starting from a small system and letting it grow in all directions until convergence is reached. The accuracy of the approximate contraction is systematically controlled by the bond dimension $\chi$ of the environment tensors. The CTMRG algorithm can also be  extended to general unit cell sizes, where separate environment tensors are kept for each position $\vec{r}$ in the unit cell, see Refs.~\cite{corboz2011, corboz14_tJ} for details.

To obtain the tensor network representation of the ground state of a given Hamiltonian, the optimal variational parameters in the tensors need to be determined. This can be achieved either by performing an imaginary time evolution~\cite{jordan2008,jiang2008,phien2015}, or by directly minimizing the energy~\cite{corboz2016, vanderstraeten2016a,liao19}. Here we use the variational optimization from Ref.~\cite{corboz2016} to obtain the initial ground states for the time evolution.

\subsection{Real-time evolution of a state} 
\label{sec:FU}

To perform a real-time evolution of a state we apply the full-update scheme from Ref.~\cite{czarnik19} (which is also used for  imaginary time evolution~\cite{jordan2008,corboz2010,lubasch2014a,phien2015})). It is based on a Trotter-Suzuki decomposition of the time-evolution operator
\begin{equation}
    \label{eq:trotter_square}
        e^{-i \ham t} = \left(e^{-i \ham dt}\right) ^k \approx  \left( \prod_{\langle m, n\rangle} \hat{g}_{mn} \right)^k
\end{equation}

\noindent where $dt = t / k$ is a small time-step and $\hat{\mc{H}}=\sum_{\langle m,n \rangle} \hat{h}_{mn}$ a Hamiltonian with nearest-neighbor terms $\hat{h}_{mn}$. Each two-body time evolution gate $\hat{g}_{mn }=e^{-i \hat{h}_{mn} dt}$ can be written as a product of two order-three tensors, with $\tilde{D}$ the bond dimension of the common index. Contracting  a single gate $\hat g_{mn}$ with neighboring tensors $\mbf{A}$ and $\mbf{B}$ of some initial state $\ket{\Psi}$ with bond dimension $D$, results in tensors $\tilde{\mbf{A}}$ and $\tilde{\mbf{B}}$ with bond dimension $D \times \tilde{D}$ in between them. Thus, the exact contraction of \cref{eq:trotter_square} with $\ket{\Psi}$ results in a  time-evolved state with bond dimension growing exponentially with $k$.

 To avoid the exponential growth, one approximates the tensors $\tilde{\mbf{A}}$, $\tilde{\mbf{B}}$ by new tensors $\mbf{A}^\prime$, $\mbf{B}^\prime$ of dimension~$D$. Let $\ket{\tilde{\Psi}} = \hat{g}_{mn} \ket{\Psi}$ and $\ket{\Psi^\prime}$ the network where the two modified tensors $\tilde{\mbf{A}}$, $\tilde{\mbf{B}}$ are replaced by the (to be found) approximations $\mbf{A}^\prime$, $\mbf{B}^\prime$. In the full-update approach this is achieved by the following minimization,
\begin{align}
    \label{eq:update_min}
    \min_{\mbf{A}^\prime, \mbf{B}^\prime} \lVert \ket{\tilde{\Psi}} - \ket{\Psi^\prime} \rVert = \min_{\mbf{A}^\prime, \mbf{B}^\prime} ( &\braket{\tilde{\Psi}}{\tilde{\Psi}}  +  \braket{\Psi^\prime}{\Psi^\prime} \nonumber \\
    &- \braket{\tilde{\Psi}}{\Psi^\prime} - \braket{\Psi^\prime}{\tilde{\Psi}} ),
\end{align}
where the CTMRG algorithm is used to compute the environment of the pair of tensors $\mbf{A'}$ and $\mbf{B'}$, based on which the four overlaps can be evaluated~\cite{corboz2010,phien2015}.
In principle, for every gate application, a new set of environment tensors would need to be recalculated from scratch. However, as the change in the state is only small after an update step, the previous environment tensors can be recycled, such that in most cases  doing a single renormalization step of the CTMRG suffices to update the environment tensors. This variant is known as the fast-full update due to its lower computational cost, and we refer to Ref.~\cite{phien2015} for  the details.

We note that, as an alternative to the full update, one may use the simple update~\cite{jiang2008} or cluster update~\cite{wang2011} optimization. These methods are typically computationally cheaper, because they do not require a full contraction of the network, but also less accurate than the (fast-) full update. These cheaper updates are particularly beneficial in the case of imaginary-time evolution-based ground state calculations, as the network only needs to be contracted once in the large imaginary time $\beta$ limit. However, in the present context the situation is different, because observables need to be evaluated at each time step, i.e. a contraction is anyway needed. Therefore,  there is no significant computational benefit in using the cheaper updates.

\begin{figure}[t]    
    \includegraphics[width=0.9\linewidth]{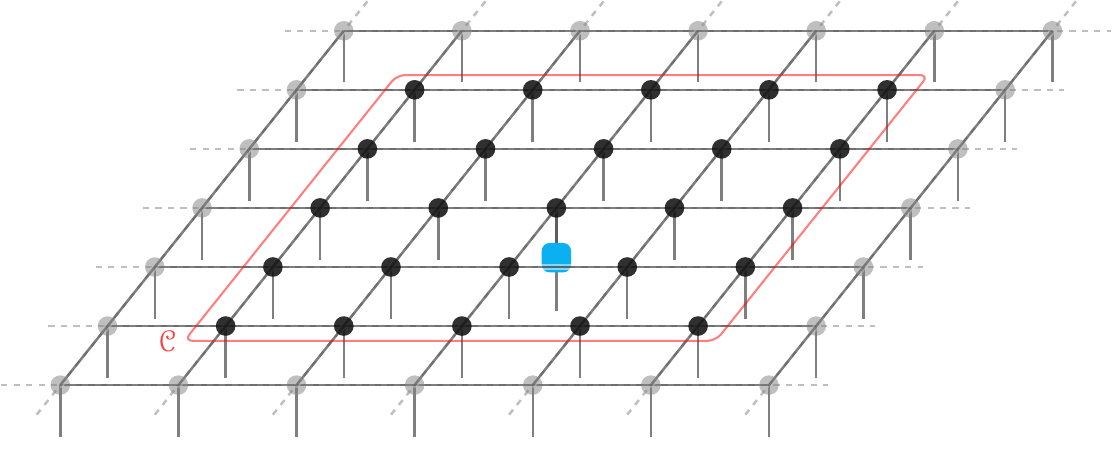}
    \caption{iPEPS representation of $\ket{\Psi(t=0)}=\hat{S}_0^x\ket{\Psi_0}$ which is used as an initial state for the time evolution. Starting from a translational invariant iPEPS ground state $\ket{\Psi_0}$, a square unit cell  $\ms{C}$ of size $L\times L$ is defined, with the operator $\hat{S}_0^x$ (light-blue) applied in the center. The unit cell is periodically repeated and chosen sufficiently large, such that  $\hat{S}_0^x$ in one  cell does not affect the neighboring cells.}
    %
    \label{fig:env_lquench}
\end{figure}

\subsection{Finite-cell dynamical iPEPS method}\label{sec:DSF}

In this section we introduce the main approach of this work for the computation of spectral functions, based on the existing real-time evolution and CTMRG algorithm discussed in the previous sections. The spectral function we  consider in this work is the dynamical structure factor (DSF) of a spin system with Hamiltonian $\ham$

\begin{equation}
    S^{\alpha, \alpha}(\vec{k}; \omega) = \frac{1}{2\pi} \int^{\infty}_{-\infty} \dt e^{i\omega t}  \expval{\hat{S}_{\vec{k}}^\alpha(t) \hat{S}_{-\vec{k}}^\alpha(0)}_c
    \label{eq:spec_fun}
\end{equation}

\noindent
defined in terms of plane waves with momentum $\vec{k}$, $\hat{S}_{\vec{k}}^\alpha(t) = \frac{1}{\sqrt{N}} \sum_{\vec{r}}e^{-i \vec{k} \cdot \vec{r}} \hat{S}_{\vec{r}}^\alpha(t)$, with $\hat{S}_{\vec{r}}^\alpha(t)= e^{i \hat{\mc{H}} t} \hat{S}_{\vec{r}}^\alpha e^{-i \hat{\mc{H}t}}$ the time-dependent local spin-operators at site $\vec{r}$ ($\alpha \in \{x,y,z\}$, $\hat{S}^\alpha_{\vec{r}} = \tfrac{1}{2} \hat{\sigma}^\alpha_{\vec{r}}$), and $\expval{\dots}_c$ indicates the connected correlation function with respect to the ground state $\ket{\Psi_0}$.   For the sake of concreteness we focus on $\alpha=x$ in the following.

To calculate the spectral function within the scope of our method, we rewrite \cref{eq:spec_fun} as

\begin{align}
    S^{x, x}(\vec{k}; \omega) &= \frac{1}{2\pi} \sum_{\vec{r}} \int^{\infty}_{-\infty} \dt e^{i(\omega t - \vec{k} \cdot \vec{r})} \expval{\hat{S}_{\vec{r}}^x(t) \hat{S}_{0}^x (0)}_c \nonumber \\
    &\approx \frac{1}{\pi} \sum_{\vec{r}} \int^{t_\mt{max}}_{0} \dt e^{-i\vec{k} \cdot \vec{r}} \; \mathrm{Re}\;\left[ e^{i \omega t}\expval{\hat{S}_{\vec{r}}^x(t) \hat{S}_{0}^x (0)}_c \right]
    \label{eq:spec_fun_ft}
\end{align}
\noindent
where in the second line we consider the case where the two-point correlator is known only for a finite time interval $\left[0, t_\mt{max}\right]$. Expressing the two point correlator as

\begin{equation}
    \expval{\hat{S}_{\vec{r}}^x(t) \hat{S}_{0}^x (0)} = e^{i E_0 t} \expval{\hat{S}_{\vec{r}}^x e^{-i \hat{\mc{H}t}} \hat{S}_0^x}{\Psi_0},
    \label{eq:2_corr_phase}
\end{equation}

\noindent
highlights the sequence of steps needed for its calculation. First, we start from tensor network representation of the ground state $\ket{\Psi_0}$ which in this work we obtain using variational optimization~\cite{corboz2016} based on a translational invariant ($L=1$) iPEPS. Next, we apply the operator $\hat{S}^x_0$, however, this breaks the translation invariance of the ground state, in other words, an exact representation of $\hat{S}_0^x\ket{\Psi_0}$ would require an infinite size unit cell. Instead, we use an iPEPS with a large square unit cell of size $L$, with $\hat{S}^x_0$ located at the center of $\ms{C}$, repeated infinitely on the lattice (\cref{fig:env_lquench}). This approximation is justified as long as the correlation length $\xi$ is much smaller than cell size, i.e. $L \gg \xi$, as in this case the effect of a $\hat{S}^x_0$ within its unit cell does not affect neighboring unit cells. That is, the perturbation on the ground state of individual copies of the operator can be treated as uncorrelated.

This representation of $\hat{S}_0^x\ket{\Psi_0} \coloneqq \ket{\Psi(t=0)}$ allows us to use existing methods to perform the time evolution and to contract the tensor network using CTMRG.
To calculate the time-evolved state $\ket{\Psi(t)} \coloneqq e^{-i \mc{H}t} \ket{\Psi(t=0)}$ we employ the fast-full update scheme, where we use larger bond dimensions of the iPEPS ($D_t$) and environment tensors ($\chi_t$) than those of the ground state, to cope with the growth of entanglement during the time-evolution. 
In practice, the time-evolution is reliable up to a time $t_\mt{max}$ after which stability issues affect the results (see Sec.~\ref{sec:entanglement_quench}), and our assumption of uncorrelated perturbations is no longer valid.
 
 To evaluate Eq.~\ref{eq:2_corr_phase} at each time step, we rewrite it in the following way, also taking into account that $|\Psi_0\rangle$ is in general not normalized:
\begin{eqnarray}
 \expval{\hat{S}_{\vec{r}}^x(t) \hat{S}_{0}^x (0)}&=&     \frac{ e^{i E_0 t} \bra{\Psi_0} \hat{S}_{\vec{r}}^x \ket{\Psi(t)}}{\bra{\Psi_0}\ket{\Psi(t)}} \frac{\bra{\Psi_0}\ket{\Psi(t)}}{\bra{\Psi_0}\ket{\Psi_0}},\nonumber \\
 &=& \frac{\bra{\Psi_0} \hat{S}_{\vec{r}}^x \ket{\Psi(t)}}{\bra{\Psi_0}\ket{\Psi(t)}}      \frac{\bra{\Psi_0} \hat{S}_0^x\ket{\Psi_0}}{\bra{\Psi_0}\ket{\Psi_0}}.    
 \label{eq:rewriteSS}
\end{eqnarray}
The first term can be computed by contracting the tensor network representing the overlap $\bra{\Psi_0}\ket{\Psi(t)}$ using CTMRG, and evaluating the local observable $\hat{S}_{\vec{r}}^x$ at each position $\vec{r}$ in the unit cell. The second term in Eq.~\ref{eq:rewriteSS}  simply corresponds to the ground state expectation value of $ \hat{S}_0^x$.

Finally, to reduce nonphysical signatures in the spectral function due to the finite time window, we multiply the time dependence of the correlator with a Gaussian envelope of the form $\exp(-\alpha \left(\frac{t}{t_\mt{max}}\right)^2)$ when transforming to frequency space.
We also define the normalized real quantity $\tilde{S}^{x, x}(\vec{k}; \omega) = \mathrm{Re}\; \left[S^{x, x}(\vec{k}; \omega)/\int \mt{d}\omega\;  S^{x, x}(\vec{k}; \omega)\right]$.

We note that the spectral function may also be computed based on finite PEPS. The main advantage of using iPEPS instead  is that it allows one to obtain results directly in the thermodynamic limit, thereby avoiding boundary and finite-size effects. Another advantage is that  one can  start from the iPEPS ground state tensor, which is just a single tensor for a translationally invariant state. For a finite PEPS with open boundaries, one would first need to find the $L\times L$ different ground state tensors, which is typically more challenging than optimizing a single tensor in the iPEPS case~\footnote{A single-tensor ansatz could be used for the ground state on periodic finite systems; however, the contraction of periodic systems is substantially harder than of open boundary systems}.

\section{Results}\label{sec:results}

\begin{figure}
    \subfloat[]{
        \includegraphics[width=0.45\textwidth]{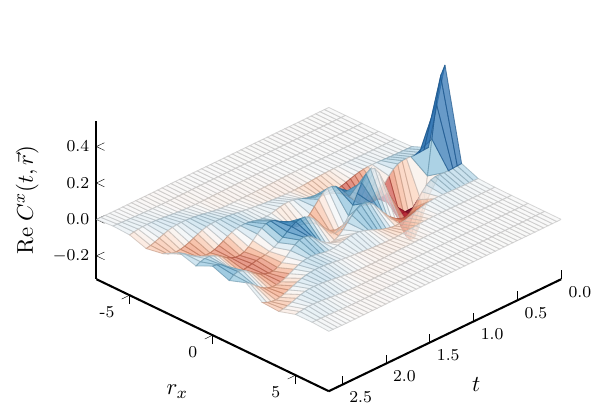}
        }

    \subfloat[]{
        \includegraphics[width=0.45\textwidth]{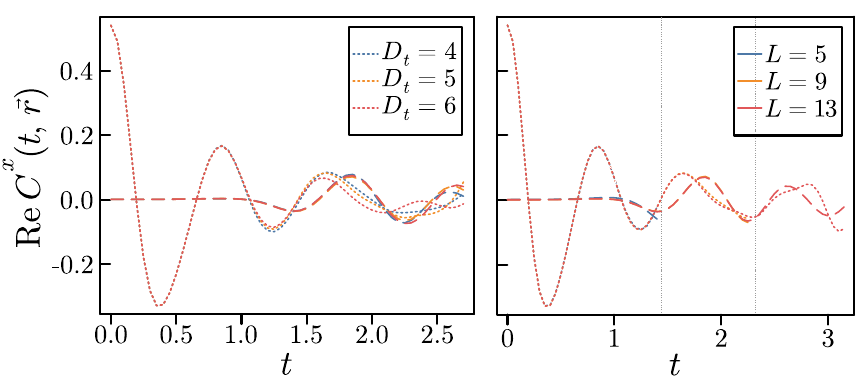}
        }
    \caption{(a) Non-equal time two-point correlator along a cross-section in the unit cell ($\vec{r}=(r_x, 0)$) for $L=15$, $\lambda=2.5$ and $D_t=5$. (b) Convergence of non-equal time two-point correlators (for $\lambda=2.5$) at two locations $\vec{r}=(0,0)$ (dotted) and $\vec{r}=(2,2)$ (dashed) as a  function of $D_t$ (left panel, $L = 11$) and  $L$ (right panel, $D_t = 5$). The vertical dotted lines indicate the maximal simulation time $t_\mt{max}$ (cf. Sec.~\ref{sec:entanglement_quench}) for $L=5$ and $L=9$, respectively.}
    \label{fig:corr_XX}
\end{figure}

\begin{figure*}
    \includegraphics[width=0.9\textwidth]{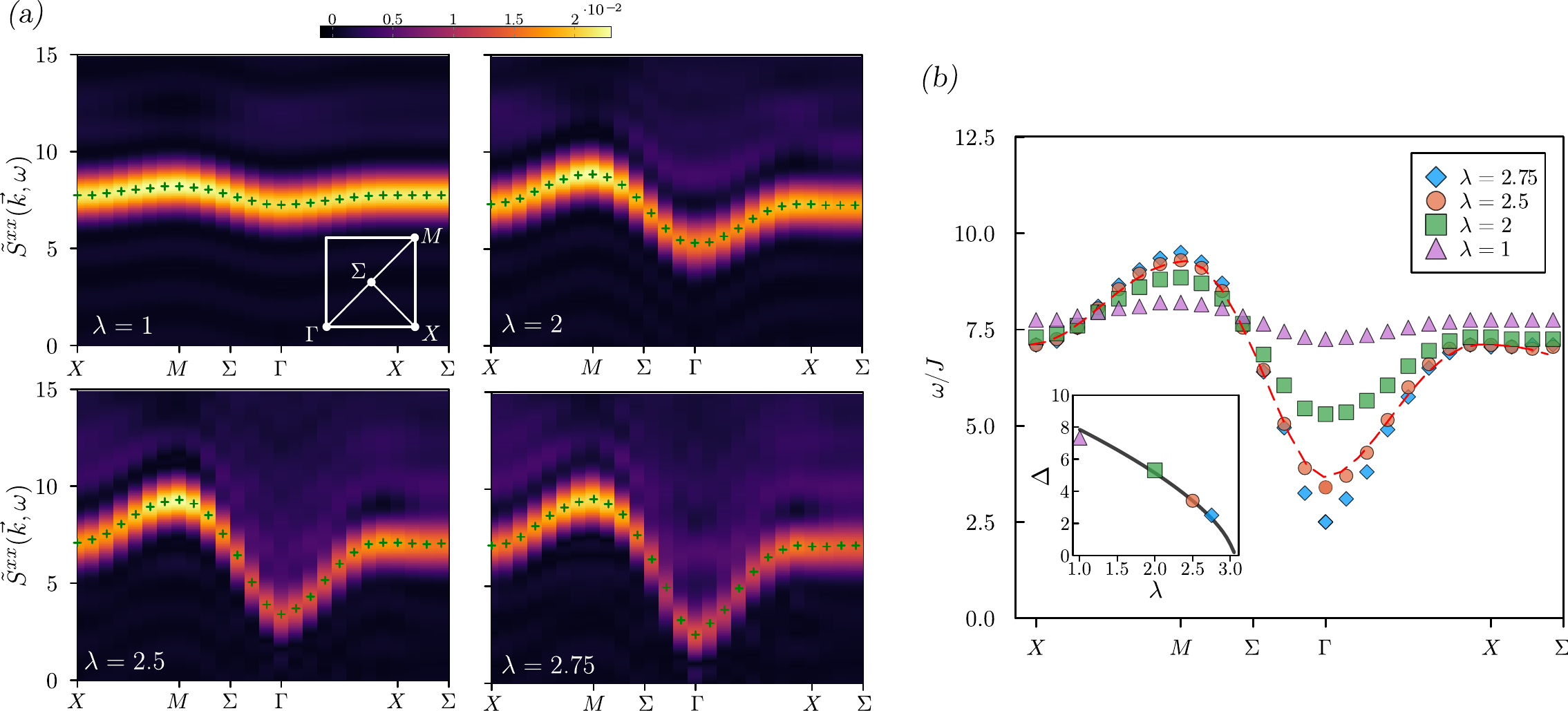}

    \caption{(a) Dynamical structure factor, $\tilde{S}^{xx}(\vec{k}, \omega)$ for $\lambda = \left\{1, 2, 2.5, 2.75\right\}$, $L=15$, and $t_\mt{max}=2.65$. The location of the maximal spectral weight at each momentum, corresponding to the magnon, is indicated by green crosses. A second (weak) feature due to the bound state can be observed (upon zooming) at higher energies. (b) Magnon dispersion curve obtained from the location of the maxima in the dynamical structure factor (\cref{fig:dsf_all}). The dashed curve is the result obtained with the iPEPS excitation ansatz~\cite{ponsioen20}. Inset: spectral gap, $\Delta$, as a function of $\lambda$, obtained at the $\Gamma$ point, where the excitation energy has a minimum. }  
    \label{fig:dsf_all}
\end{figure*}

As a benchmark of our method we consider the ferromagnetic transverse-field 2D Ising model on a square lattice  
\begin{equation}
    \hat{\mc{H}} = - J \sum_{\langle i,j \rangle}  \hat{\sigma}^z_i \hat{\sigma}^z_j - \lambda \sum_{i } \hat{\sigma}^x_i,
\end{equation}
with  $J$ nearest-neighbor coupling and $\lambda$ the transverse magnetic field strength. The model exhibits a second-order phase transition between a ferromagnetic phase and a disordered phase at $\lambda_c/J \approx 3.044$~\cite{blote2002}. In the following we set $J=1$.

For all our simulations, we use a ground state optimized for bond dimensions of $D_0=3$ and  $\chi_0 = 40$. We chose a time step $dt=0.05$, which is small enough such that the Trotter error remains small. 
For the time-evolution, we use  bond dimensions in the range of $D_t= 3-7$ and a sufficiently large $\chi_t = 50 - 100$, such that contraction errors are small. Finally, for the Gaussian smoothing of the correlator, $\alpha$ is fixed to $2.0$.

\subsection{Local quench dynamics}\label{sec:quench}

We first present example results for the non-equal two-point correlator $  C^x(t,\vec{r}) = \expval{\hat{\sigma}_{\vec{r}}^x(t) \hat{\sigma}_{0}^x (0)}_{c} $ based on a $L=15$ unit cell and $\lambda=2.5$ in \cref{fig:corr_XX}(a), from where the propagation of the perturbation $\hat{S}_0^x$ on the ground state can be observed. As for any local system, the correlations (carried by quasiparticles) propagate with a finite velocity until reaching the boundary of the unit cell. Consequently, the finite cell size  constrains the simulation times, as will be discussed in Sec.~\ref{sec:entanglement_quench}.

The accuracy of the time evolution is controlled on the one hand by the finite bond dimension $D_t$, restricting the amount of entanglement the ansatz can reproduce, and the finite unit cell size $L$.  In \cref{fig:corr_XX}(b) we study the effect of $D_t$ and  $L$ on the non-equal time correlator for two different distances $\vec{r}$ and $\lambda=2.5$. We find that finite $D_t$ effects are small up to $t\sim 1.5$, beyond which they become visible in the longer-distance correlator. 

Regarding the dependence on $L$, small differences can be observed only when approaching the maximum simulation times (cf. Sec.~\ref{sec:entanglement_quench}) for each~$L$.

\subsection{Dynamical structure factor}

Based on the data for the non-equal time correlators we calculate the dynamical structure factor for four values of $\lambda$ in the ferromagnetic phase ($\lambda = 1, 2, 2.5, 2.75$) for a path through the Brillouin zone as shown in \cref{fig:dsf_all}. We can clearly identify a sharp feature with the largest spectral weight, corresponding to the single-particle excitation (magnon) of the system, which has energy $8J$ in the classical $\lambda=0$ limit. An additional distinct feature at higher energies is observed, particularly for $\lambda\le2$, which can be associated with a bound state of two magnons, corresponding to a pair of spin-flips on neighboring sites. In the limit $\lambda=0$ it has an energy of $12J$ as opposed to $16J$ of two free magnons. For $\lambda>2$ this feature becomes less distinct, in particular around the $\Gamma$ point, and a broader region with enhanced spectral weight can be identified, compatible with a continuum of two-magnon excitations.

\begin{figure}
    \centering
    \includegraphics[width=0.45\textwidth]{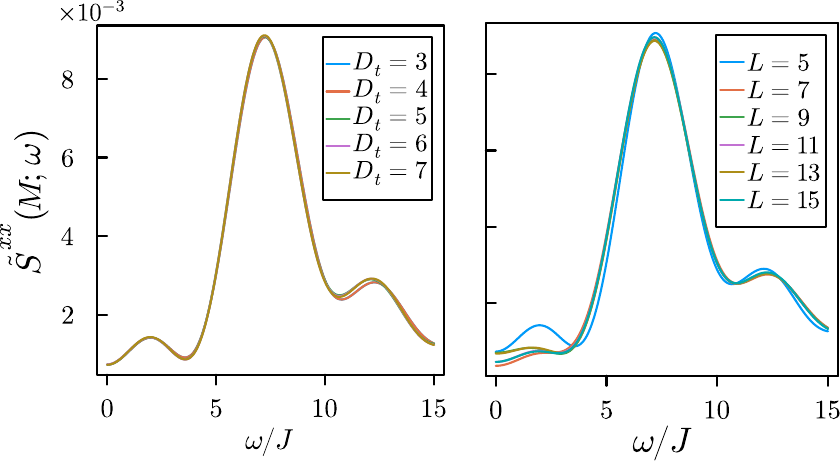}
    \caption[]{Effect of $D_t$ (left, fixed $L=5$) and $L$ (right, fixed $D_t=5$) on dynamical structure factor at the $M=(\pi,\pi)$ point for $\lambda=2.5$. All results are obtained up to a $t_\mt{max}=1.4$. }
    \label{fig:Sqw_Dt}
\end{figure}

In \cref{fig:Sqw_Dt} we show  data for the dependence of the dynamical structure factor on $D_t$ and $L$ for $\lambda=2.5$ at momentum $\vec{k}=(\pi,\pi)$ as an example. The finite $D_t$ and $L$ effects are found to be rather weak, except that for the smallest cell size $L=5$ an artificial small peak appears at low energy, which vanishes for the larger cell sizes. Thus, already relatively small values of $D_t$ and $L$ are sufficient here to reproduce the main features in the dynamical structure factor. Stronger effects may be found for longer time scales $t_\mt{max}$, which, however, is limited by the cell size, as we discuss in the next section.

From the location of the maximum in the dynamical structure factor for each value of $\vec{k}$, we extract the magnon dispersion curves plotted in \cref{fig:dsf_all}(b). For $\lambda = 2.5$ we also present a comparison with results obtained with the iPEPS excitation ansatz~\cite{ponsioen20}, showing good agreement with our results, apart from a slight deviation at the $\Gamma$ point. We can also extract the spectral gap from the values at momentum $\vec{k}=(0,0)$ ($\Gamma$ point), plotted in \cref{fig:dsf_all}(b). Its dependence on $\lambda$ follows closely the expected scaling law $\Delta \propto \abs{\lambda - \lambda_c}^\nu$, with $\nu = 0.63012$~\cite{campostrini2002}. Deviations are expected when approaching the critical point $\lambda_c$ as the finite bond dimension in the ground state induces an effective finite correlation length at criticality~\cite{corboz18,rader18}.

\begin{figure}
    \centering
    \includegraphics[width=0.45\textwidth]{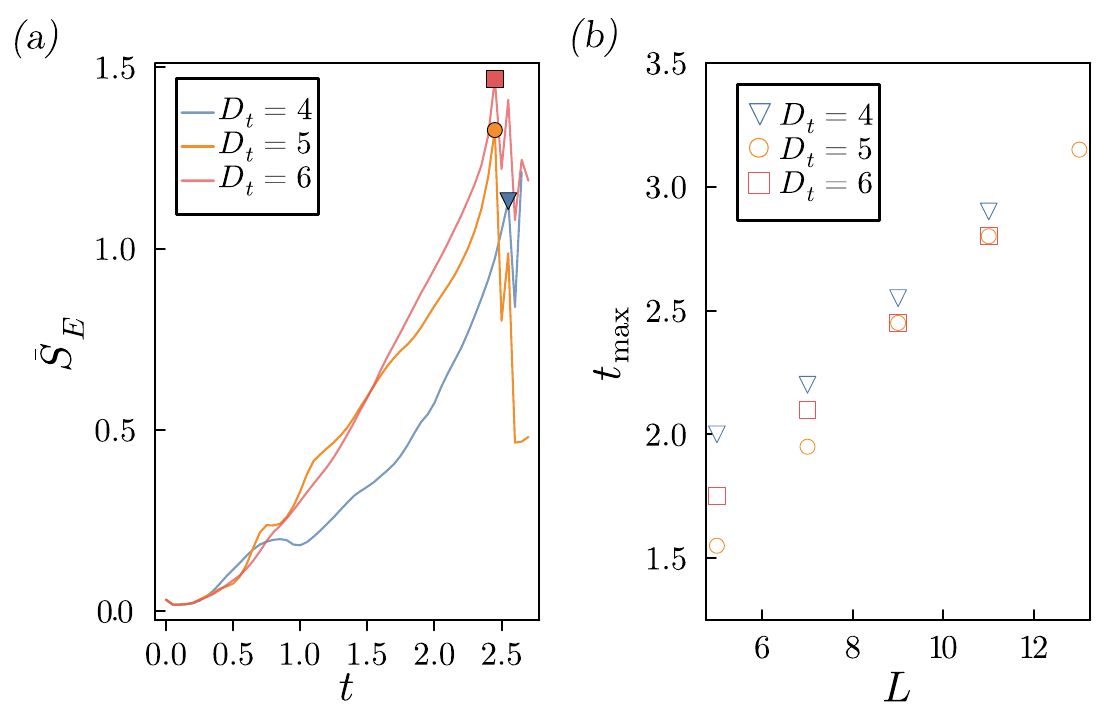}
    \caption{(a) Average value of $S_E$ for $L=9$. (b) Maximum simulation times $t_\mt{max}$ as a function of cell size $L$ for different bond dimensions $D_t$, extracted from the discontinuity in the growth of $\bar{S}_E$. $\lambda=2.5$ in both figures.}
    \label{fig:t_max}
\end{figure}

\subsection{Maximal simulation times}\label{sec:entanglement_quench}

As seen in the previous sections, it is possible to reproduce the main features of the non-equal time correlation functions and dynamical structure factor for moderate values of $D_t$, indicating that the entanglement growth during the time evolution can be sufficiently captured with our ansatz up to the $t_\mt{max}$ considered. Thus, in the present simulations, the finite $D_t$ is not the limiting factor on the maximal simulation times, but rather the cell size $L$, as we discuss in the following. 

It is clear that when the propagating correlations from the initial perturbation reach the boundary of the cell, they are going to be influenced by the propagating correlations from the neighboring cell, such that the assumption of uncorrelated cells is no longer valid. What is more, we observe that after a characteristic time when the correlations reach the boundary, the time evolution becomes numerically unstable, which manifests itself by an abrupt change of observables.

For example, we can asses the stability of the simulations by monitoring the growth of entanglement through the quantity  $S_E = -\sum_j \eta_j \log \eta_j$, obtained from the eigenvalues $\left\{\eta_j\right\}$ of a corner tensor of the environment, and its mean value, $\bar{S}_E$, given by the average over all corner tensors. This quantity has the form of a von Neumann entropy and we expect it to increase with increasing entanglement entropy of an infinite quadrant of the wavefunction.

In \cref{fig:t_max}(a) we observe that after a certain time the growth of $\bar{S}_E$ exhibits a sudden change and we identify the last time step before the instability occurs  as the maximum simulation time $t_\mt{max}$ up to which we can extract reliable data. We find that $t_\mt{max}$ increases roughly linearly with cell size $L$, as shown in \cref{fig:t_max}(b), suggesting that $t_\mt{max}$ can be related to the characteristic time scale when the propagating correlations reach the boundary of the cell. The dependence of $t_\mt{max}$ on $D_t$ is found to be rather weak for the larger system sizes. 

Thus, we conclude that in order to reach longer time scales, the cell size $L$ has to be increased, which comes with a computational cost that scales as ${\cal O}(L^2)$. 

\subsection{Complementarity of methods}\label{sec:entanglement_quench}
As mentioned in the introduction, the dynamical structure factor can also be obtained based on the iPEPS excitation ansatz~\cite{vanderstraeten2015}. This ansatz is based on the ground state iPEPS with a local perturbation, i.e. where one of the ground state tensors is replaced by a new tensor $B$,
\begin{equation}
  \ket{\Psi_0} \to \ket{\Phi(B)_{\bm{\vec{r}}}}.
\end{equation}
By taking an infinite sum over all positions of $B$ with a appropriate phase factors, the ansatz exhibits a well defined momentum by construction,
\begin{equation}
  \ket{\Phi(B)_{\bm{k}}} = \sum_{\bm{\vec{r}}} e^{i \bm{k} \cdot \bm{\vec{r}}} \ket{\Phi(B)_{\bm{\vec{r}}}}.
\end{equation}
Several methods to optimize this ansatz have been developed; see Refs.~\cite{ponsioen20,ponsioen22,ponsioen23,tu24} for details. From the set of all excitations, the dynamical structure factor can then be obtained via a spectral decomposition of $S(\vec{k}; \omega)$. 

The real-time evolution based method and the excitation ansatz provide two complementary tools to study spectral functions, each with its own strengths and weaknesses. 
Due to the nature of the excitation ansatz, it can resolve single quasiparticle-like excitations (or bound states) with high accuracy, especially also at low energies. 
However, a disadvantage is that it can only approximately resolve a continuum of excitations, which may result in spurious sharp spectral features in the dynamical structure factor instead of a clear continuum at small D~\footnote{This limitation may be partially overcome by a two-particle excitation ansatz, which has been tested for MPS~\cite{vanderstraeten2014}, but has not yet been generalized to iPEPS.}.

In contrast, the time-evolution-based approach is not inherently biased towards single-particle excitations, allowing it to more naturally resolve a continuum of excitations. Another advantage is that a single calculation provides the full dynamical structure factor for all momenta, whereas a separate calculation is needed for each value of $k$ for the excitation ansatz. However, to reach the same level of accuracy for single-particle excitations at low energies as the excitation ansatz, progress on the time-scales reachable by the time-evolution method is needed.

\section{Conclusion}\label{sec:outlook}
In this work, we have shown how current iPEPS methods for infinite 2D quantum systems can be combined to evaluate non-equal time correlators to compute spectral functions. Our approach is based on an iPEPS ansatz consisting of copies of large unit cells with an operator applied in the center of each cell, which is time-evolved using the fast full-update, and 2-point correlators are evaluated with CTMRG.  We have presented benchmark results for the transverse field Ising model, and found that the main features in the dynamical structure factor can be reproduced already at relatively small bond dimensions $D_t$ and cell size $L$.  

We obtained results for the magnon dispersion which are in good agreement with the ones based on the iPEPS excitation ansatz~\cite{ponsioen20}. Due to the nature of the ansatz, the latter is more accurate to represent single-particle excitations (or bound states), however,  it can capture multiparticle continua only in a limited way, in contrast to time-evolution based methods. Thus, the two approaches provide complementary tools to study spectral functions.

In future it would be desirable to push the simulations to longer times to improve the resolution of spectral features. In the present approach we found that the maximal simulation time is mostly limited by the finite cell size, and related to the characteristic time when correlation effects between neighboring cells become non-negligible, leading to instabilities in the time evolution. While longer time scales may be obtained using larger cell sizes, another possible direction could be to use a modified ansatz consisting of only  a single cell with an operator applied, surrounded by an effective environment of the unperturbed state. The challenge is then to time evolve the latter alongside with the iPEPS tensors in the cell. This could potentially be achieved based on a single-layer CTMRG scheme~\cite{lubasch2014a, pizorn2011}, which keeps separate bra- and ket- environment tensors with an effective physical index in between.  A similar idea has  successfully been tested for infinite MPS in Ref.~\cite{phien2012}. Other interesting directions include the study of real-time scattering of interacting quasiparticles in 2D~\cite{vandamme2021a, vanderstraeten2015a}, which could potentially be done based on the present real-time evolution approach, and the extension to time-evolutions of a perturbation in momentum space~\cite{vandamme2022}.

\acknowledgments
We acknowledge discussions with R. Gerritsma and A. Safavi-Naini. This project has received funding from the European Research Council (ERC) under the European Union's Horizon 2020 research and innovation programme (grant agreement Nos. 677061 and 101001604). This work is part of the D-ITP consortium, a program of the Netherlands Organization for Scientific Research (NWO) that is funded by the Dutch Ministry of Education, Culture, and Science (OCW).

\bibliography{refs_peps,refs}

\end{document}